\documentstyle[12pt,epsfig]{article}
\newcommand{\beq}{\begin{equation}}
\newcommand{\eeq}{\end{equation}}
\newcommand{\bea}{\begin{eqnarray}}
\newcommand{\eea}{\end{eqnarray}}

\newcommand{\gsim}{\lower.7ex\hbox{$
\;\stackrel{\textstyle>}{\sim}\;$}}
\newcommand{\lsim}{\lower.7ex\hbox{$
\;\stackrel{\textstyle<}{\sim}\;$}}

\newcommand{\eod}{\end{document}}

\def\ot{{\bf T}}
\def\cp{{\bf CP}}
\def\cpt{{\bf CPT}}

\newcommand{\NP}{New Physics}

\begin{document}
\thispagestyle{empty}
\vspace*{-22mm}

\begin{flushright}
UND-HEP-08-BIG\hspace*{.08em}03\\
CAS-KITPC/ITP-056 

\end{flushright}
\vspace*{1.3mm}

\begin{center}
{\LARGE{\bf
"I Know She Invented Fire, but What Has She Done Recently?"\\ 
or\\
"I Have Come to Praise Ch., not Bury Her!" \\
or\\
`On the Second Renaissance of Charm Physics' \\
}}
\footnote{Lecture given at "2nd Workshop on Theory, Phenomenology and 
Experiments in Heavy Flavour Physics', June 16 - 18, 2008, Capri, Italy}
\vspace*{19mm}

{\Large{\bf I.I.~Bigi}} \\
\vspace{7mm}

{\sl Department of Physics, University of Notre Dame du Lac}
\vspace*{-.8mm}\\
{\sl Notre Dame, IN 46556, USA}\\
{\sl email: ibigi@nd.edu}

\vspace*{10mm}

{\bf Abstract}\vspace*{-1.5mm}\\
\end{center}

\noindent
$D^0$ oscillations, for which the $B$ factories have found strong evidence, provide 
a new stage for our search for New Physics in heavy flavour dynamics. While the theoretical verdict on the observed values of $x_D$ and $y_D$ is ambiguous -- they could be fully generated by SM dynamics, yet could contain also a sizable contribution from New Physics -- such oscillations 
can enhance the observability of \cp~violation driven by New Physics. After emphasizing the unique role of charm among up-type quarks, I sketch the \cp~phenomenology for partial widths and final state distributions.

\tableofcontents

\vspace{1.5cm}

While the first title reflects the spontaneous reaction of a large part of the HEP community, when they 
hear about charm physics, the second one conveys the message I want to communicate to you, which is condensed into a more conventional form in the third title. The formulation 
"Second Renaissance" makes reference to the "First  Renaissance" discussed this morning, which was prompted by the surprises in the spectroscopy of hadrons with open and hidden charm. 

\begin{figure}[ht]
\begin{center}
\epsfig{bbllx=-10cm,bblly=13cm,bburx=20cm,bbury=23cm,
height=6truecm, width=18truecm,
        figure=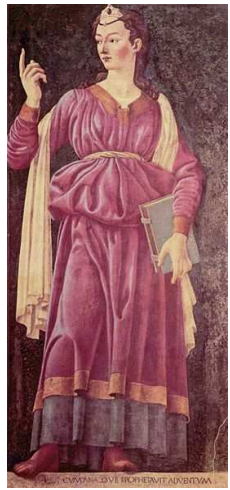}
\caption{Painting of the Cumean Sybil by Andrea del Castagno
\label{SYBIL}  
}
\end{center}
\end{figure}

Yet first I want to remark on the `genius loci' of Capri that might not be well known to non-Italians. The most venerated oracle in ancient Italy was that of Cumae near Naples with the `ageless' Cumaean 
Sybil (or prophetess) presiding over it as priestess. A portrait by is shown in Fig.\ref{SYBIL}
This allows me to address two points relevant for our meeting: 

\noindent (i) Giulia thus stands in a long tradition of female bosses in this part of the world. Keep that in mind when even considering disagreeing with her. 

\noindent 
(ii) There is an intriguing legend about the Cumaean Sybil. She had offered nine books with all her prophecies to the last Roman king Tarquinius Superbus  for sale. Considering the asking price to stiff, he declined. She then threw three of the books into a fire to burn them and asked the same price for the remaining six books. He still refused, whereupon she burnt three more books. Then he relented and bought the left over three books for the original asking price.The experimentalists among you will recognize that Tarquinius Superbus acted like the typical funding agency that asks for `de-scoping' your project only to end up paying the same price for less. 
The theorists will claim that if we had nine flavours to study, we would already have figured out the dynamics underlying the flavour enigma \footnote{As pointed out by P. Colangelo in his talk there is 
a difference between `enigma' and `puzzle'.}. 

Back to the main subject. 
While the study of strange dynamics was instrumental for the creation of the 
Standard Model (SM) and that of charm transitions central for it being accepted, the analysis of $B$ decays almost completed its validation through the establishment of CKM dynamics as the dominant  
source of the observed \cp~violation; `almost', since the Higgs boson has not been observed yet. Now the race is on to see which of these areas together with top quark decays will reveal an incompleteness of the SM in flavour dynamics. If the evidence for 
$D^0$ oscillations with $x_D$, $y_D \sim 0.005 - 0.01$ gets confirmed, then the detailed probe of \cp~symmetry in charm decays is just behind the race leader, namely the even more detailed study of  $B$ decays. 

The signal for $D^0$ oscillations marks a {\em tactical} draw: while the values measured for 
$x_D$ and $y_D$ might be generated by SM forces alone, they could contain relatively large contributions from \NP~(NP). Yet a {\em strategic} victory is in sight: studies of 
\cp~symmetry in $D$ decays will decide the issue possibly paving the way for a {\em new} SM 
to emerge. I would like to draw a historical analogy based on my personal experience. Sanda and myself had been talking about large 
\cp~asymmetries in $B$ decays \cite{BS80} 
without much resonance -- till $B_d - \bar B_d$ oscillations were resolved by the ARGUS collaboration in 1987 \cite{ARGUS87}, i.e. twenty-one years ago. 
Yet quantitatively we have a `centi-ARGUS' scenario with the oscillation parameter $x_D$ being about two orders of magnitude smaller than $x_B$. 
\cp~asymmetries in $D$ decays will be smaller than what was found in $B$ decays. However the `background' from SM dynamics is even tinier. I would also count on our experimentalists having become more experienced and thus being able to extract smaller signals. 

The outline of the talk is as follows: after a Prologue on the unique place of charm studies in searches 
for New Physics I review our inconclusive interpretation of the data on $D^0$ oscillations before my 
central message -- the need for a comprehensive search for \cp~violation in charm decays. After an Outlook I conclude with an Epilogue on the shift between `Capri I' and `Capri II'.  

\section{Prologue: On Charm's Unique Place}

NP in general induces flavour changing neutral currents (FCNC). The SM and many viable NP models had to be carefully crafted to suppress FCNC below acceptable levels. 
FCNC could be much stronger for {\em up}-type than for down-type quarks -- quite unlike the 
situation within the SM. This actually happens in some models which `brush the dirt' of FCNC in the down-type sector under the `rug' of the up-type sector. 

With the SM `background' smaller for FCNC of up-type quarks, we can hope for cleaner though not 
necessarily larger NP signals there: 
\beq
\left. \frac{\rm NP \, signal}{\rm SM \, `noise'}\right|_{\rm up-type} > 
\left. \frac{\rm NP \, signal}{\rm SM \, `noise'}\right|_{\rm down-type} 
\eeq

{\em Among up-type quarks it is only charm that allows the full range of probes for 
FCNC and New Physics in general}: (i) Top quarks decay {\em before} they can hadronize 
\cite{RAPALLO}. Without top {\em hadrons} 
$T^0 - \bar T^0$ oscillations cannot occur. This limits our options 
to search for \cp~asymmetries, since one cannot call on oscillations to provide the required second amplitude. 
(i) Hadrons built with $u$ and $\bar u$ quarks like $\pi^0$ and $\eta$ are their own antiparticle; thus there can be no $\pi^0 - \pi^0$ etc. oscillations as a matter of principle. Furthermore they possess so few 
decay channels that \cpt~invariance basically rules out  \cp~asymmetries in their decays.  

I will show that only recently have experiments reached a range of sensitivity, where one can realistically expect 
\cp~violation to show up in charm transitions. My basic contention is as follows: {\em Charm transitions are a unique portal for obtaining novel access to flavour dynamics with the experimental situation being a priori favourable apart from the absence of Cabibbo suppression.} 

\section{On the Interpretation of $D^0$ Oscillations}

In the limit of \cp~invariance oscillations are described by the normalized mass and width 
splittings:  $x_D \equiv \frac{\Delta M_D}{\Gamma_D}$, 
$y_D \equiv \frac{\Delta \Gamma_D}{2\Gamma_D}$. 
While the SM predicts similar numbers for $x_D$ and $y_D$ with the data showing the same trend, we should note that $\Delta M_D$ and $\Delta \Gamma_D$ reflect different dynamics:  
$\Delta M_D$ is produced by {\em off}-shell transitions making it naturally sensitive to NP  
unlike $\Delta \Gamma_D$, which is generated by {\em on}-shell modes.  A central 
theoretical issue is to which degree quark-hadron duality can be invoked, in particular for 
$\Delta \Gamma_D$, which involves less averaging or `smearing' than $\Delta M_D$; or 
in more general terms: how sensitive is $\Delta \Gamma_D$ to the proximity of several 
hadronic thresholds \cite{CICERONE}.

\subsection{Theoretical Estimates and Data}

Within the SM two reasons combine to make $x_D$ and $y_D$  
small in contrast to the situation for $B^0 - \bar B^0$ and  
$K^0 - \bar K^0$ oscillations, namely the double Cabibbo suppression of the 
amplitude for $D^0 \leftrightarrow \bar D^0$ coupled with the GIM suppression being controlled 
by the breaking of $SU(3)_{fl}$. A rather conservative bound reads \cite{CICERONE}: 
\beq
x_D, \, y_D \sim {\rm SU(3)_{fl}\, break.} \times 2{\rm sin}^2\theta_C < {\rm few}\times 0.01
\eeq
The description of $SU(3)_{fl}$ breaking becomes a  
central issue. While $x_D \ll y_D$ would be 
unnatural, it cannot be ruled out.  
The history of the predictions on $D^0$ oscillations does not provide a tale of 
consistently sound judgment by theorists, when they predicted 
$x_D \leq {\rm few}\times 10^{-4}$. Yet 
scientific progress is not made by majority vote, although that codifies it in the end. 
It should be noted that words of caution had been sounded; e.g. in 1997 
\cite{VARENNA97}: {\em "... It is often stated that the SM predicts ... 
$x_D, y_D \leq 3\cdot 10^{-4}$. I myself am somewhat flabbergasted by the boldness of such 
predictions ... I cannot see how anyone can make such a claim with the required confidence ... "} Warnings similar in substance -- albeit more 
diplomatic in tone -- had been sounded by Wolfenstein and Donoghue. 

In estimating the strength of ${\cal L}(\Delta C=2)$ authors had typically relied on evaluating quark box diagrams that had been faithful guides for ${\cal L}(\Delta S=2)$ and ${\cal L}(\Delta B=2)$, 
while overlooking the fact that the resulting GIM suppression of $(m_s/m_c)^4$ is un-typically severe. 
The often heard statement that oscillations of mesons built from {\em up} type quarks teach us about {\em down} type quark dynamics -- which is inspired by looking at quark box diagrams with charged currents -- is thus misleading. The correct statement is that those oscillations tell us about the FCNC of 
{\em up} type quarks. 

Two complementary approaches to evaluating $\Delta M_D$ and $\Delta \Gamma_D$ in the SM represent the state of the art. They can be referred to as `fully inclusive' and 
`summing over exclusive channels'. 

\noindent 
In the `inclusive' approach one constructs an operator product expansion (OPE) in terms of operators 
constructed from quark and gluon fields and takes their expectation value. There is one new element 
relative to what has been done with great success in $B$ decays: One has to include contributions 
from quark condensates -- i.e., vacuum expectation values $\langle 0|\bar qq|0\rangle$ -- in addition to 
$D^0$ expectation values, since the quark box contributions, which represent the partonic term, are so severely suppressed here, as mentioned above. Thus one has to deal with three  
parameters with mass dimension, namely $m_c$, $m_s$ and the condensate scale $\mu _{had}$.  Since  
$\mu _{had}$ and $m_c$ are comparable in size, the resulting OPE is not a very robust one, at least 
numerically. One finds that the largest contribution is ${\cal O}(m_s^2\mu _{had}^4/m_c^6)$ 
rather than the formally leading quark box term ${\cal O}(m_s^4/m_c^4)$ \cite{CICERONE}: 
\beq 
x_D(SM)|_{OPE}, \, y_D(SM)|_{OPE} \sim {\cal O}(10^{-3})
\eeq
with a slight preference for $x_D(SM)|_{OPE} < y_D(SM)|_{OPE}$; their relative sign is 
not predicted. On the other hand one infers
\beq 
{\rm arg}M_{12}^D(SM)/\Gamma_{12}^D(SM)|_{OPE} \sim 
\lambda ^4 \eta \leq 10^{-3}
\eeq
As stated before, violations of quark-hadron duality due to the relative proximity of several relevant 
production thresholds could enhance in particular $y_D$ over this estimate. In any case 
it appears quite unlikely that the theoretical uncertainty of this estimate can be reduced. 

\noindent 
The other approach \cite{FALK1} operates on the purely hadronic rather than quark-gluon level. 
2-, 3- and 4-body modes are considered with $SU(3)_{fl}$ breaking in the decay rates identified 
with that due to their phase space alone. Summing over these groups of channels  yields an estimate for 
$\Delta \Gamma_D$ and a dispersion relation for $\Delta M_D$: 
\beq 
y_D(SM) \sim 0.01 \; , \; \; 0.001 \leq |x_D(SM)| \leq y_D 
\eeq
with $y_D(SM)$ and $x_D(SM)$ being of opposite sign; arg$M_{12}^D/\Gamma_{12}^D$ cannot be predicted this way. 

In evaluating the theoretical situation one has to distinguish carefully between two similar sounding 
questions: 
\begin{enumerate}
\item 
What is the most likely SM value for $x_D$, $y_D$? My answer is as before: ${\cal O}(10^{-3})$. 
\item 
Can one rule out $0.01$? There I say `no'!

\end{enumerate}

In the Spring of 2007 intriguing evidence for $D^0$ oscillations has been presented 
by the BaBar \cite{BABAROSC} and Belle \cite{BELLEOSC} collaborations. Averaging over them 
HFAG \cite{HFAG} finds $(x_D,y_D) \neq (0,0)$ with more than 6 sigma significance; more 
specifically: 
\beq 
x_D = (0.89^{+0.26}_{-0.27})\% \; , \; \; y_D = (0.75^{+0.17}_{-0.18})\%
\eeq
I fervently hope that more precise measurements will confirm these oscillation signals with 
$x_D$ and $y_D$ in the range 0.5 - 1\%. Establishing $D^0$ oscillations would provide a novel insight into flavour dynamics. After having discovered oscillations in {\em all three} mesons built from 
{\em down}-type quarks -- $K^0$, $B_d$ and $B_s$ -- it would be the first observation of oscillations with 
{\em up}-type quarks; it would also remain the only one (at least for three-family scenarios), as explained above. 

\subsection{Interpretation?}

It would have been conceivable to measure $y_D \ll x_D \sim {\rm few}\times 0.01$ thus establishing 
the intervention of NP. This has not happened: we are in a grey zone, where the observed strengths of both $y_D$ and $x_D$ might be produced by SM forces alone -- or could contain significant contributions from NP. Even in the former case one should probe 
these oscillations as accurately as possible first establishing $[x_D,y_D] \neq [0,0]$ and then determining $x_D$ vs. $y_D$. Analogous to the situation with 
$\epsilon^{\prime}/\epsilon_K$ one has to aim at measuring them irrespective of limitations in 
our theoretical tools. 

A future theoretical breakthrough might allow us to predict $x_D|_{SM}$ and 
$y_D|_{SM}$ 
more accurately and thus resolve the ambiguity in our interpretation, but I would not count on it. Rather than wait for that to happen the community should become active in the catholic tradition of `active repentance' and search for \cp~violation in $D$ decays. Even if NP is not the main engine for  
$\Delta M_D$, it could well be the leading source of \cp~violation in ${\cal L}(\Delta C=2)$. 
There is an analogy to the case of $B_s$ oscillations. 
$\Delta M (B_s)$ has been observed to be 
consistent with the SM prediction within mainly theoretical uncertainties; yet since those are 
still sizable, we cannot rule out that NP impacts $B_s$ oscillations 
significantly. This issue, which is unlikely to be resolved soon theoretically, can be decided experimentally 
by searching for a time dependent \cp~violation in $B_s(t) \to \psi \phi$. For within the SM one predicts \cite{BS80} a very small asymmetry not exceeding 4\% in this transition since on the leading CKM level quarks of only the second and third family contribute. Yet in general one can expect NP contributions to $B_s$ oscillations to exhibit a weak phase that is not particularly 
suppressed. Even if NP affects $\Delta M(B_s)$ only moderately, it could greatly enhance 
the time dependent \cp~asymmetry in $B_s(t) \to \psi \phi$. 
This analogy is of course qualitative rather than quantitative with 
$D^0$ oscillations being quite slow.

\section{\cp~Violation -- the Decisive Stage}

\subsection{On NP Effects}

Probing \cp~invariance for manifestations of NP is not a `wild goose chase'. For we know that CKM dynamics is completely irrelevant for baryogenesis; i.e., we need \cp~violating NP to understand the Universe's observed baryon number as a {\em dynamically generated} quantity rather than an arbitrary initial value. 

There is no need to construct crazy NP scenarios for charm transitions -- being innovative will do. At present we have the "usual list of suspects" \cite{CASA}: {\em Non-}minimal SUSY with(out) R parity 
(up-squarks might be less degenerate than down-squarks), Higgs dynamics with{\em out} 
natural flavour conservation, Little Higgs models, extra dimensions etc. I do not know of  
persuasive NP scenarios that would affect $D$ decays, but not $B$ and $K$ decays. Yet their 
manifestations might stand out more clearly in $D$ where there is little SM `background'. 
It behooves us to show some humility in judging whether a scenario is persuasive. For while 
we know so much about flavour dynamics, we understand very little. Probing \cp~symmetry in 
charm transitions is certainly of the `hypothesis-generating' rather than `hypothesis-probing' variety.

Charm decays offer several {\em pragmatic} advantages in such searches: 

\noindent
(i) While we do not know how to reliably compute the strong phase shifts 
required for direct \cp~violation to emerge in partial widths, we can expect them to be large, 
since charm decays proceed in an environment populated by many resonances. 
{\em Hadronization thus 
enhances the observability of \cp~violation}; it `only' causes a problem when we attempt to 
interpret the findings in terms of microscopic NP parameters. 

\noindent 
(ii) The branching ratios into relevant modes 
are relatively large. 

\noindent 
(iii) Asymmetries being linear in NP amplitudes enjoy enhanced sensitivity to the latter. 

\noindent 
(iv) The soft pions from $D^* \to D\pi$ provide a powerful tagging tool. 

\noindent 
(v) Many $D$ decays lead to three or more pseudoscalar mesons with various 
resonant structures. This complexity allows \cp~asymmetries to surface in final state distributions rather than merely in partial widths, and significantly larger asymmetries might arise in the former than the latter. 

\noindent 
(vi)  The `background' from known physics is small. According to the SM there is a three-level Cabibbo hierarchy with the rates of Cabibbo allowed, once and doubly Cabibbo suppressed modes scaling roughly like $1:1/20:1/400$. The SM makes non-trivial predictions for each of these Cabibbo levels. 
{\em Without} oscillations direct \cp~violation can 
arise only in {\em singly} Cabibbo suppressed transitions, where it is driven by the highly diluted 
phase $\sim \lambda^4 \eta$ of $V(cs)$. One expects asymmetries to reach no better than the 0.1 \% level; significantly larger values would signal NP. 
{\em Almost any} asymmetry in Cabibbo 
{\em allowed} or {\em doubly suppressed} channels requires the intervention of New Physics, since -- 
in the absence of oscillations -- there is only one weak amplitude. The exception are channels containing 
a $K_S$ (or $K_L$) in the final state like $D \to K_S \pi$. There are two sources for a 
\cp~asymmetry from known dynamics: (i) Two transition amplitudes are actually involved,  
a Cabibbo favoured and a doubly suppressed one, $D \to \bar K^0\pi$ and $D \to K^0\pi$, respectively. 
Their relative {\em weak} CKM phase is given by $\eta A^2 \lambda ^6 \sim {\rm few} \cdot 10^{-5}$, which seems to be well beyond observability. (ii) While one has $|T(D \to \bar K^0 \pi)| = 
|T(\bar D \to K^0 \pi)|$, the \cp~impurity $|p|\neq |q|$ in the $K_S$ wave function introduces a difference 
between $D^{0,+}\to K_S\pi^{0,+}$ and $\bar D^{0,-}\bar K_S \pi^{0,-}$ of 
$\frac{|q|^2 - |p|^2}{|q|^2 + |p|^2} = (3.32 \pm 0.06)\cdot 10^{-3}$ \cite{CICERONE}.

With oscillations on an observable level -- and it seems $x_D$, $y_D$ $\sim 0.005 - 0.01$ satisfy 
this requirement -- the possibilities for \cp~asymmetries proliferate. Those will allow us to decide whether NP is involved.

\subsection{Oscillations as New \cp~Portal} 

In the presence of $D^0 - \bar D^0$ oscillations 
{\em time-dependent} \cp~asymmetries 
can arise in $D^0$ decays on the Cabibbo allowed 
($D^0 \to K_S\phi$ 
\footnote{Since the final state $K_S\phi$ is mainly given by a single isospin amplitude, the 
strong phase basically drops out from 
$T(\bar D^0 \to K_S\phi)/T(D^0 \to K_S\phi)$; i.e., the \cp~asymmetry measures the 
NP weak phase.} , $K_S\rho^0$, $K_S\pi ^0$), 
once forbidden ($D^0 \to K^+K^-$, $\pi^+\pi^-$) and doubly forbidden $(D^0 \to K^+\pi^-$) levels. Let me list just two prominent 
examples from the last two categories. Since $y_D$, $x_D \ll 1$, it suffices to give the 
decay rate evolution to first order in those quantities only (the general expressions can be found in 
Ref.\cite{CICERONE}). 
\bea 
\nonumber
\Gamma (D^0(t) \to K^+K^-) &\propto &  e^{-\Gamma_1t}|T(D^0 \to K^+K^-)|^2 \times  
\\
&&\left[ 1 +y_D\frac{t}{\tau_D} ( 1 - {\rm Re}\frac{q}{p}\bar \rho_{K^+K^-} ) - 
x_D\frac{t}{\tau_D}{\rm Im}\frac{q}{p}\bar \rho_{K^+K^-}\right] \\
\nonumber 
\Gamma (\bar D^0(t) \to K^+K^-) &\propto & e^{-\Gamma_1t}|T(\bar D^0 \to K^+K^-)|^2\times 
\\
&&\left[ 1 +y_D\frac{t}{\tau_D} ( 1 - {\rm Re}\frac{p}{q} \frac{1}{\rho_{K^+K^-}} ) - 
x_D\frac{t}{\tau_D}{\rm Im}\frac{p}{q}\frac{1}{\rho_{K^+K^-}} \right]
\label{DKK} 
\eea
The usual three types of \cp~violation can arise, namely the direct and indirect types -- 
$|\bar \rho_{K^+K^-}| \neq 0$ and $|q|\neq |p|$, respectively -- as well as the one involving 
the interference between the oscillation and direct decay amplitudes -- 
Im$\frac{q}{p}\bar \rho_{K^+K^-}\neq 0$ leading also to Re$\frac{q}{p}\bar \rho_{K^+K^-}\neq 1$. 
Assuming for simplicity \footnote{CKM dynamics is expected 
to induce an asymmetry not exceeding 0.1\%.} $|T(D^0 \to K^+K^-)| = |T(\bar D^0 \to K^+K^-)|$ 
and $|q/p| = 1- \epsilon_D$ with $|\epsilon_D| \ll 1$ one has 
$(q/p)\bar \rho_{K^+K^-} = (1-\epsilon_D) e^{i\phi_{K\bar K}}$ and thus 
\beq
 \frac{\Gamma (\bar D^0(t) \to K^+K^-) - \Gamma (D^0(t) \to K^+K^-)}
{\Gamma (\bar D^0(t) \to K^+K^-) + \Gamma (D^0(t) \to K^+K^-)} 
\simeq x_D\frac{t}{\tau_D} {\rm sin}\phi_{K\bar K} -  
y_D\frac{t}{\tau_D}\epsilon_D {\rm cos}\phi_{K\bar K}\; .  
\eeq
BELLE has found \cite{BELLEOSC} for such an asymmetry integrated over time:
\beq 
A_{\Gamma} = (0.01 \pm 0.30 \pm 0.15) \%
\eeq
While there is no evidence for \cp~violation in the transition, one should also note that 
the asymmetry is bounded by $x_D$, $y_D$. For $x_D$, $y_D \leq 0.01$, as indicated by the data, 
$A_{\Gamma}$ could hardly exceed the 1\% range; i.e., there is not much of a 
bound on $\phi_D$ or $\epsilon_D$ so far. Yet any improvement in the experimental sensitivity for  
$D^0(t) \to K^+K^-$ constrains NP scenarios -- or could reveal them 
\cite{GKN}. 

Another promising channel for probing \cp~symmetry is $D^0(t) \to K^+\pi^-$: since it is 
doubly Cabibbo suppressed, it should a priori exhibit a higher sensitivity to a 
New Physics amplitude.  Furthermore it cannot exhibit direct \cp~violation in the SM. 
With 
\beq 
\frac{q}{p} \frac{T(D^0 \to K^+\pi^-)}{T(D^0 \to K^-\pi^+)}
\left[   \frac{p}{q} \frac{T(\bar D^0 \to K^-\pi^+)}{T(\bar D^0 \to K^+\pi^-)}  \right] \equiv  
 - \frac{1}{{\rm tg}^2\theta_C} (1-[+] \epsilon_D) |\hat \rho _{K\pi}|e^{-i(\delta - [+]\phi_{K\pi})}
\eeq
one expresses an asymmetry as follows: 
$$  
\frac{\Gamma (\bar D^0(t) \to K^-\pi^+) - \Gamma (D^0(t) \to K^+\pi^-)}
{\Gamma (\bar D^0(t) \to K^-\pi^+) + \Gamma (D^0(t) \to K^+\pi^-)} \simeq \; \; \; \; 
$$
\beq 
 \left(\frac{t}{\tau_D}\right) \left| \hat \rho_{K\pi}\right|
 \left( \frac{y_D^{\prime}{\rm cos}\phi_{K\pi}\epsilon_D - 
x_D^{\prime}{\rm sin}\phi_{K\pi}}{{\rm tg}\theta_C^2}\right) + 
 \left(\frac{t}{\tau_D}\right)^2 \left| \hat \rho_{K\pi}\right|^2 \frac{\epsilon_D(x_D^2 + y_D^2)}
 {2{\rm tg}\theta_C^4}
 \eeq
 where I have again assumed for simplicity $|\epsilon _D| \ll 1$ and {\em no direct} 
 \cp~violation. 
 
BABAR has searched for a time dependent \cp~asymmetry in $D^0 \to K^+\pi^-$ vs. 
$\bar D^0(t) \to K^- \pi^+$, yet so far not found any evidence \cite{BABAROSC}. Again, with $x_D^{\prime}$ and $y_D^{\prime}$ capped by 
about 1\%, no nontrivial bound can be placed on the weak phase $\phi_{K\pi}$. On the other hand any further increase in experimental sensitivity could reveal a signal. 

\subsection{On \cpt~Constraints}

\cpt~symmetry provides more constraints than just equality of mass and lifetime of particles and antiparticles. For it tells us that the widths for subclasses of transitions have to be the same. For 
simplicity consider a toy model where the $D$ meson can decay only into two classes of final states 
$A=\{ a_i, i=1,...,n \}$ and $B=\{b_j, j=1,...,m \}$ with the strong interactions allowing members of the 
class $A$ to rescatter into each other and likewise for class $B$, but {\em no} rescattering possible 
{\em between} classes $A$ and $B$. Then \cpt~symmetry tells us partial width asymmetries 
{\em summed} over class $A$ already have to vanish and likewise for class $B$. This \cpt~`filter' can hardly be of any 
practical use for $B$ decays with their multitude of channels, yet for $D$ decays it might provide nontrivial validation checks. Details can be found in Ref.\cite{CPBOOK}.

\subsection{Final State Distributions}

Decays to final states of {\em more than} two pseudoscalar or one pseudoscalar and one vector meson contain 
more dynamical information than given by their  widths; their distributions as described by Dalitz plots 
or \ot{\em -odd} moments can exhibit \cp~asymmetries that can be considerably larger than those for the 
width. All \cp~asymmetries observed so far in $K_L$ and $B_d$ decays 
except one concern partial widths, i.e. 
$\Gamma (P \to f)$ $\neq$ $\Gamma (\bar P \to \bar f)$. The one 
notable exception can teach us important lessons for future searches both in charm and $B$ decays, namely the \ot~odd moment found in $K_L\to \pi ^+ \pi ^- e^+ e^-$. Denoting by $\phi$ the angle between the $\pi^+\pi^-$ and $e^+e^-$ planes 
one has 
\beq 
\frac{d\Gamma}{d\phi}(K_L \to \pi^+\pi^- e^+e^-) =  \Gamma_1 {\rm cos}^2 \phi + 
\Gamma_2 {\rm sin}^2 \phi + \Gamma_3 {\rm cos} \phi {\rm sin}\phi 
\eeq
Comparing the $\phi$ distribution integrated over two quadrants one obtains a 
\ot~{\em odd} moment: 
\beq 
\langle A \rangle = \frac{\int _0^{\pi/2}d\phi \frac{d\Gamma}{d\phi} - \int _{\pi/2}^{\pi}d\phi \frac{d\Gamma}{d\phi}}
{\int _0^{\pi}d\phi \frac{d\Gamma}{d\phi}}= \frac{2\Gamma_3}{\pi (\Gamma_1 + \Gamma_2)}
\label{<A>}
\eeq
$\langle A \rangle $ is measured to be $0.137 \pm 0.015$ \cite{PDG06} in full agreement with the prediction of 
$0.143 \pm 0.013$ \cite{SEGHALKL}. 
Most remarkably this large asymmetry is generated by the tiny \cp~impurity parameter $\eta_{+-}\simeq 0.0024$; i.e., the impact 
of the latter is magnified by a factor of almost a hundred -- for the price of a tiny branching ratio of about $3\cdot 10^{-7}$! 

Likewise one might find larger \cp~asymmetries in final state distributions of three-, four-body etc. 
$D$ decays like $D\to 3\pi$, $K \bar K\pi$, $K\bar K \pi \pi$, $K\bar K\mu^+\mu^-$. As far as 
three-body modes are concerned we have a `catholic' scenario: there is a single canonical path 
to heaven -- the Dalitz plot. Four-body modes on the other hand represent  a `Calvinist scenario': 
while {\em a priori} many paths can lead to heaven -- generalized Dalitz studies, angular asymmetries in the decay planes as sketched above for $K_L \to \pi^+\pi^- e^+e^-$ etc. -- Heaven's blessing will be revealed {\em a posteriori 
through success}. A pilot study of 
$D^0 \to K^+K^-\pi^+\pi^-$ vs. $\bar D^0 \to K^+K^-\pi^+\pi^-$ has been undertaken by the 
FOCUS collaboration \cite{PEDRINI}.

\subsection{Semileptonic $D^0$ Decays}

$|q/p| \neq 1$ unambiguously reflects \cp~violation in $\Delta C=2$ dynamics. It can be probed most directly in semileptonic $D^0$ decays leading to `wrong sign' leptons: 
\beq
a_{SL}(D^0) \equiv  \frac{\Gamma (D^0(t) \to l^-X) - \Gamma (\bar D^0 \to l^+X)}
{\Gamma (D^0(t) \to l^-X) + \Gamma (\bar D^0 \to l^+X)} = 
\frac{|q|^4 - |p|^4}{|q|^4 + |p|^4} 
\eeq
The corresponding observable has been studied in semileptonic decays of neutral $K$ and $B$ mesons. With $a_{SL}$ being controlled by $(\Delta \Gamma/\Delta M){\rm sin}\phi_{weak}$, 
it is predicted to be small in both cases, albeit for different reasons: 
(i) While $(\Delta \Gamma_K/\Delta M_K) \sim 1$ one has sin$\phi_{weak}^K \ll 1$ leading to 
$a_{SL}^K = \delta _l  \simeq (3.32 \pm 0.06)\cdot 10^{-3}$ as observed. 
(ii) For $B^0$ on the other hand one has 
$(\Delta \Gamma_B/\Delta M_B)\ll1$ leading to $a_{SL}^B < 10^{-3}$. 

For $D^0$ both $\Delta M_D$ and $\Delta \Gamma_D$ are small, yet 
$\Delta \Gamma_D/\Delta M_D$ is not: present data indicate it is about unity; 
$a_{SL}$ is given by the smaller of $\Delta \Gamma_D/\Delta M_D$ or its inverse multiplied by 
sin$\phi_{weak}^D$, which might not be that small: i.e., while the rate for `wrong-sign' leptons is small in semileptonic decays of neutral 
$D$ mesons, their \cp~asymmetry might not be at all, if New Physics intervenes to generate 
$\phi_{weak}^D$.  

\subsection{Benchmark Goals}

Viable NP scenarios could produce \cp~asymmetries close to the present experimental bounds, but 
hardly higher. To have a realistic chance to find an effect, one should strive to reach at least 

$\bullet$ 
the ${\cal O}(10^{-4})$ [${\cal O}(10^{-3})$] level for time-dependent \cp~asymmetries in 
$D^0 \to K^+K^-$, $\pi^+\pi^-$, $K_S\rho^0$, $K_S\phi$ [$D^0 \to K^+\pi^-$]; 

$\bullet$ 
direct \cp~asymmetries in partial widths down to ${\cal O}(10^{-3})$ in $D\to K_S\pi$ and  in 
singly Cabibbo suppressed modes and down to ${\cal O}(10^{-2})$ in doubly Cabibbo suppressed 
modes; 

$\bullet$
the ${\cal O}(10^{-3})$ level in Dalitz asymmetries and \ot~odd moments.

\section{Conclusions and Outlook}

It is important to firmly establish the existence of $D^0$ oscillations and determine $x_D$ vs. $y_D$. 
My main message is that we must go after \cp~violation in charm transitions in all of its possible manifestations, both 
time dependent and independent, in partial widths and final state distributions, and on all Cabibbo levels down to the 
$10^{-3}$ or even smaller level. The present absence of any \cp~asymmetry is not telling. Comprehensive and detailed \cp~studies of charm decays provide a unique window onto flavour dynamics. 

For that purpose we need the statistical muscle of LHCb. Charm studies constitute a worthy challenge to LHCb, for which 
$D^0 \to K^+K^-$, $\pi^+\pi^-$, $K^+\pi^-$, $K^+K^-\pi^+\pi^-$, $K^+K^-\mu^+\mu^-$ represent good channels. On the 
theory side we can expect a positive learning curve for theorists, yet should not count on 
miracles. 
Therefore we have to go after even more 
statistics and more channels, including those with (multi)neutrals to validate our future conclusions. 
This brings me to my second message: 
\begin{center} 
"Ceterum censeo fabricam super saporis esse faciendam!" 

"Moreover I advise a super-flavour factory has to be built!" 
\end{center}
Such a machine could provide an even more optimal environment, if it could be operated also 
at charm threshold with decent luminosity. 

\section{Epilogue: From Capri I to Capri II}
\begin{figure}[ht]
\begin{center}
\epsfig{bbllx=1.5cm,bblly=2cm,bburx=20cm,bbury=23cm,
height=11truecm, width=12truecm,
        figure=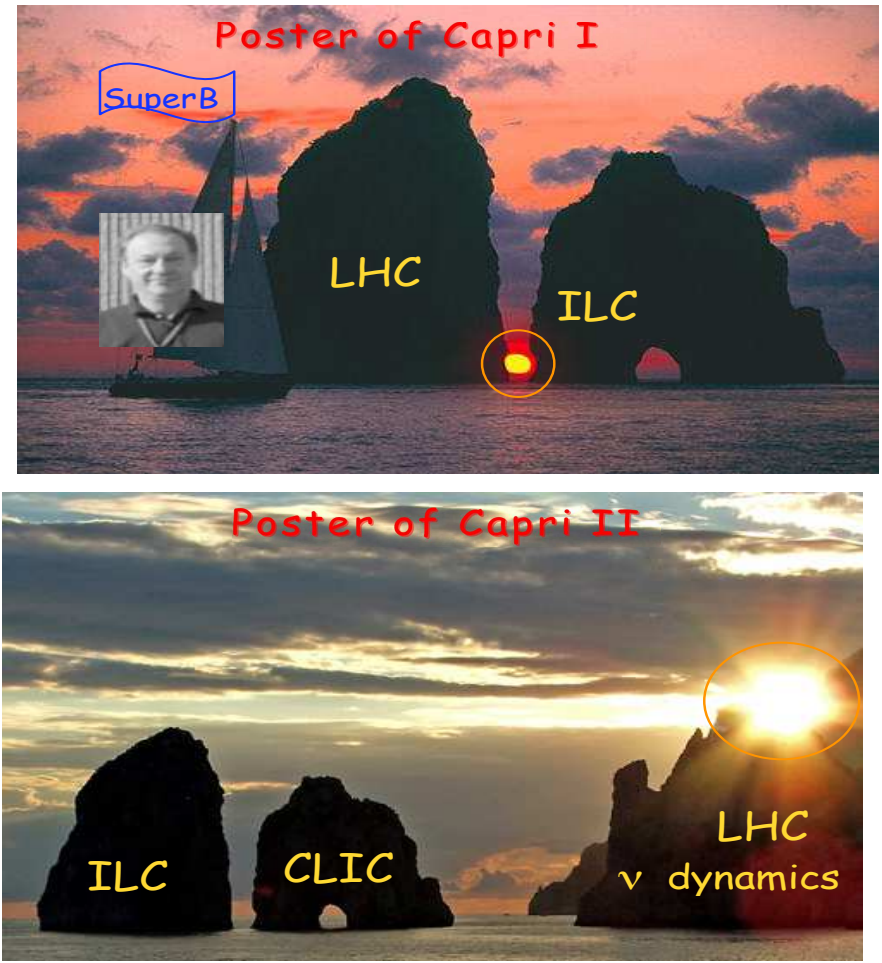}
\caption{Modified posters of the Capri I (top) and Capri II (bottom) workshops}
\label{POSTERS}  
\end{center}
\end{figure}
While I had failed to attend the first workshop here in Capri, I still used its poster for a talk at DIF06 
appropriately (as I thought) modified, see Fig.\ref{POSTERS}. Among other things it led to a question: Does it show a rising or a 
setting sun? I was quite intrigued when I saw that the poster for Capri II reflected the changed landscape of HEP in a rather poetic way (being `subtlety challenged' I have illustrated these changes, 
see Fig.\ref{POSTERS});  it contains two further messages: (i) The vitality of the light rays indicates it must be a rising sun and (ii) the passage for Super-B has become wider!

A final thought: Models with extra dimensions have several ad-hoc features. But they are 
sufficiently radical to push our thinking out of its present comfort zone into novel fruitful 
directions; i.e., they are a most helpful `imagination stretcher' in the language of L. Sehgal.


{\bf Acknowledgments:} This work was supported by the NSF under the grant number PHY-0807959. 
I am most grateful to G. Ricciardi for organizing a wonderful workshop in a most inspiring setting. 
I would like to thank also Prof. Y.-L. Wu for the the gracious hospitality extended to me at the Kavli Institute of the Chinese Academy of Science in Beijing, where this talk has been written up.

\vspace{4mm}


\end{document}